 \documentclass[12pt]{article}
   \newcommand{\bea}{\begin{eqnarray}}
 \newcommand{\eea}{\end{eqnarray}}

 \usepackage{amsfonts}
 \usepackage{amssymb}
 \usepackage[dvips]{graphicx}
\usepackage{latexsym}
 
 \newcommand{\overskrift}[1]{\vspace{3.0mm}\noindent\textbf{#1}\vspace{1.5mm}}
 \newcommand{\storoverskrift}[1]{\vspace{3.0mm}\noindent{\Large\textbf{#1}}\vspace{1.5mm}}
 
 \newcommand{\beq}{\begin{equation}}
 \newcommand{\eeq}{\end{equation}}
 \newcommand{\lpart}{\raise.3ex\hbox{$\stackrel{\leftarrow}{\partial}$}}
 \newcommand{\rpart}{\raise.3ex\hbox{$\stackrel{\rightarrow}{\partial}$}}
 \newcommand{\ldr}{\raise.3ex\hbox{$\stackrel{\leftarrow}{\delta^r}$}}

 \newcommand{\al}{\alpha}
 \newcommand{\be}{\beta}
 
 \newcommand{\p}{\partial}
\newcommand{\GeV}{~\rm GeV}
\newcommand{\MeV}{~\rm MeV}

\begin{document}

 \begin{titlepage}
 \begin{flushleft}
        \hfill                      {\tt hep-ph/0109214}\\ \hfill
        HIP-2001-51/TH \\ \hfill            September 24, 2001\\
 \end{flushleft}
 \vspace*{3mm}
 \begin{center}
 {\Large {\bf Adiabatic CMB perturbations \\} {\bf in Pre-Big-Bang string cosmology\\}}
 \vspace*{12mm} {\large Kari
 Enqvist\footnote{E-mail: kari.enqvist@helsinki.fi} and
 Martin S. Sloth\footnote{E-mail:
 martin.sloth@helsinki.fi}\\}

 \vspace{5mm}

 {
 Helsinki Institute of Physics \\ P.O. Box 9,
 FIN-00014 University of Helsinki, Finland }

 \vspace{3mm}

 and

 \vspace{3mm}

 {
 Department of Physics \\ P.O. Box 9, FIN-00014
 University of Helsinki, Finland}

 \vspace*{10mm}
 \end{center}


 \begin{abstract} \noindent
We consider the Pre-Big-Bang scenario with a massive axion field
which starts to dominate energy density when oscillating in an
instanton-induced potential and subsequently reheats the universe
as it decays into photons, thus creating adiabatic CMB
perturbations. We find that the fluctuations in the axion
field can give rise to a nearly flat spectrum of adiabatic
perturbations with a spectral tilt $\Delta n$ in the range $ -0.1 \lesssim
\Delta n \lesssim 1$.

 \end{abstract}

 \end{titlepage}

 \baselineskip16pt

 \section{Introduction}

In the Pre-Big-Bang scenario (PBB) \cite{1}, which is an attempt
to derive the standard Friedman-Robertson-Walker cosmology (FRW)
from a fundamental theory of quantum gravity, a FRW universe
emerges dynamically after an infinitely long era of dilaton
driven super-inflation. This is the so called Pre-Big-Bang era.
In the infinite past the universe approaches asymptotically a
flat, cold, and empty string perturbative vacuum state. During the
PBB era the curvature is growing until it reaches the string
scale. At this point the universe enters an intermediate string
phase where the curvature stops growing before it ultimately, by
some unknown graceful exit mechanism, ends up in the FRW era with
a decreasing curvature. The problem of the graceful
exit remains largely unsolved, although some promising
suggestions has been presented \cite{1a}.

It has been shown that the amplification of axion quantum
fluctuations during the PBB era can lead to initial isocurvature
fluctuations with a spectral tilt in an interesting range for the
CMB anisotropies \cite{2}. However, the new measurements by the
Boomerang balloon flight \cite{3} definitely show that the CMB
anisotropies are not of purely isocurvature nature \cite{4}. This
presents a very difficult hurdle for the PBB scenario and is the
main observational objection against it.

However, as was first noticed in \cite{5} in a different context,
a decaying axion field could change the situation dramatically.
If the axion field carries power at large scales, then in the FRW
era it could dominate the energy density. When it decays into
photons, instead of isocurvature fluctuations, one is lead to
initial adiabatic density perturbations in the PBB scenario \cite{12}.

It has also been noted that a periodic potential for the axion
will damp the fluctuations on large scales, avoiding eventual
divergences of the large scale axion field fluctuations
\cite{5,6,7}. The purpose of the present paper is to point out
that in the case where the axion, or one of the axions, can decay,
there exists a possibility for the PBB scenario to yield adiabatic
initial fluctuations with a nearly flat spectrum.

We will assume the generation of the axion field fluctuations
during the PBB era and that at some early point during the FRW
era the axion acquires a periodic potential due to instanton
effects \cite{8}. This happens at
 some temperature $T_{\Gamma}$ whence the axion mass starts to build up.
As soon as the axion mass is of the order of the Hubble rate
$m_a\simeq 3H$, the axion field starts oscillating and all the
long wavelength modes contribute to the total energy density.
Given a periodic potential of the type
 \beq
V(\psi)=\frac{1}{2}V_0\left(1-\cos\left(\frac{\psi}{\psi_0}\right)\right)~,
 \eeq
and assuming that the axion field is Gaussian distributed with a
zero average and a variance $\left<\psi^2(\vec{x})\right>$, the
average density is given by
 \beq \label{x}
\rho_a=\left<V(\psi(\vec{r}))\right>\approx
\frac{1}{2}V_0\left(1-e^{-\frac{1}{2}\frac{\left<\psi^2(\vec{x})\right>}{\psi_0^2}}\right)~.
 \eeq
Oscillating axions will behave like non-relativistic dust and
their energy density will grow compared to the radiation energy
density until the axion decays into photons. As we shall argue in
section IV, the axions will most likely get to dominate energy
density. The fluctuations in the axion energy density will in this
way be converted into initial adiabatic fluctuations of radiation.

We presume that we have several axions, one of which decays. In fact, in the effective 4-dimensional theory resulting from compactification of 10-dimensional string theory, one will generally end up with many axion fields. In \cite{8}, where the instanton induced axion potential was discussed, one had two axion fields: the string theory axion, which we are interested in here, and the QCD invisible axion. In \cite{8a} it was pointed out that the presence of a global $SL(n,\mathbb{R})$ symmetry of the 4-dimensional effective action is a completely general consequence of toroidal compactification from $D+n$ to $D$ dimensions. In a model with $SL(6,\mathbb{R})$ invariance, which is the maximal invariance for compactification from 10 to 4 dimensions, one would have as many as $15$ different axion fields. Cold dark matter (CDM) could for instance be the invisible QCD axion,
but it is possible that there are other stable fields with the
right properties, such as the moduli fields. We should also like to point out that in models with a $SL(4,\mathbb{R})$ symmetry or higher, there exist regions of the parameter space where the axion with the smallest spectral tilt, which will dominate over the other axions, has a flat or blue spectrum \cite{8a}.

The paper is organized as follows. In section II we will review
how the axion field fluctuations are generated in the Pre-Big-Bang phase \cite{9,10} and set up our notation. In section III we
will estimate the spectral tilt of density fluctuations induced
by the decay of the axion in the periodic potential, and in
section IV we calculate the life time of the massive axions and
discuss the reheat temperature. Section V contains our
conclusions.

\section{Initial axion field fluctuations}
Let us begin our discussion with the following four-dimensional effective
tree-level action,
 \beq
S_{eff}=\frac{1}{2\al'}\int
d^4x\sqrt{-g}e^{-\phi}\left[\mathcal{R}+\p_
{\mu}\phi\p^{\mu}\phi-\frac{1}{2}\p_{\mu}\beta
\p^{\mu}\beta+\mathcal{L}_{matter}\right]~,
 \eeq
where $\sqrt{\al'}=\sqrt{8\pi}/M_s$ and $M_s$ is the string mass.
Here $\phi$ and $\beta$ are the universal four-dimensional moduli
fields. This action can be derived from 10-dimensional string
theory compactified on some 6-dimensional compact internal space.
The matter Lagrangian, $\mathcal{L}_{matter}$, is composed of
gauge fields and axions, which we assume to be trivial constant
fields and not to contribute to the background. In the following
we will therefore only consider their quantum fluctuations.

The solution to the equations of motion for the background fields
can in the spatially flat case be parameterized as
\cite{1,9,11,12}
 \beq
a=a_s\left|\frac{\eta}{\eta_s}\right|^{(1+\sqrt{3}\cos\zeta)/2}\qquad,\qquad
e^{\phi}=e^{\phi_s}\left|\frac{\eta}{\eta_s}\right|^{\sqrt{3}\cos\zeta}\qquad,\qquad
e^{\beta}=e^{\beta_s}\left|\frac{\eta}{\eta_s}\right|^{\sqrt{3}\sin\zeta}~.
 \eeq

Let us split out the part of the matter Lagrangian that contains
the axion field
$\mathcal{L}_{matter}=\mathcal{L}_{gauge}+\mathcal{L}_{axion}$,
where we will assume that the axion Lagrangian has the general
form
 \beq
\int d^4 x\sqrt{-g}e^{-\phi}\mathcal{L}_{axion}\propto \int d\eta
d^3 x a^{2}e^{l\phi}e^{m\be}(-\sigma'^2+(\nabla \sigma)^2)~,
 \eeq
where $'$ denotes $\p/\p\eta$ and $a$ is the scale factor of the
four-dimensional metric $g_{\mu\nu}=a^2(\eta)diag(-+++)$. Note
that for the model independent axion $m=0$, $l=1$ while for the
Ramond-Ramond axion $m\neq 0$, $l=0$.

Let us for simplicity consider a model with two cosmological
phases
 \beq
\bar{a} =\bar{a}_s \left(\frac{-\eta}{\eta_s}\right)^{\al}~~
,\qquad \eta<-\eta_s
 \eeq
 \beq
\bar{a} =\bar{a}_s \left(\frac{\eta}{\eta_s}\right)~~ ,\qquad
\eta>\eta_s
 \eeq
where $\bar{a}$ is the scale factor in the axion frame, defined by
 \beq
\bar{a}\equiv e^{l\phi/2}e^{m\be/2}a
 \eeq
and $a$ is the scale factor in the string frame. We assume that
the dilaton $\phi$ and the moduli $\be$ are frozen in the post
Big Bang era. The axion frame is given by a conformal rescaling
of the string frame $g_{\mu\nu}\to\Omega^2g_{\mu\nu}$,
$\Omega=e^{l\phi/2}e^{m\be/2}$ \cite{9,11,12}. In the axion frame
the axions are minimally coupled and the evolution equation for
the canonical normalized axion field fluctuation. By matching
the PBB solution and its derivative to the FRW post Big Bang
solution at $\eta=\eta_s$, one finds that during the post Big
Bang era \cite{9},
 \beq \label{fieldosc}
\delta\sigma=\frac{C(r)}{\bar{a}\sqrt{2k}}|k\eta_s|^{-r-1/2}\sin(k\eta)~,
 \eeq
where $r=|\al-1/2|$ and
 \beq
|C(r)|=\frac{2^r\Gamma(r)}{2^{3/2}\Gamma(3/2)}~.
 \eeq

Outside the horizon $k\eta<1$, in the post Big Bang era, one can
approximate the sine function by its argument and we get
 \beq
\delta\sigma=\frac{C(r)}{\sqrt{2}}e^{-l\phi_s/2}e^{-m\be_s/2}\frac{\sqrt{\eta_s}}{a_s}|k\eta_s|^{-r}~.
 \eeq
where $a_s$ is the scale factor in the string frame at the
transition. In the post Big Bang era one assumes
$\phi\equiv\phi_s$, $\be\equiv \be_s$. Now let us define the
r.m.s. amplitude of the perturbation in a logarithmic $k$
interval $\delta\psi_k\equiv k^{3/2}\delta\sigma$ as in \cite{6}.
With this definition one obtains
 \beq
\delta\psi_k=\frac{C(r)}{\sqrt{2}}e^{-l\phi_s/2}e^{-m\be_s/2}M_s|k\eta_s|^{3/2-r}~.
 \eeq
Above we used that $a_s\eta_s=a_s/k_s=1/M_s$.

Inside the horizon the field fluctuations oscillates in time, and
one can read of the r.m.s. amplitude $A$ from Eq. (\ref{fieldosc})
 \beq
A=\frac{C(r)}{2}\frac{1}{\bar{a}\sqrt{k}}|k\eta_s|^{-r-1/2}=\frac{C(r)}{2}\frac{\sqrt{\eta_s}}{\bar{a}}|k\eta_s|^{-r-1}~.
 \eeq
Defining inside the horizon $\delta\psi_k\equiv k^{3/2}A$, we get
in agreement with \cite{6},
 \beq
\delta\psi_k=\frac{C(r)}{2}e^{-l\phi_s/2}e^{-m\be_s/2}M_s|k\eta_s|^{1/2-r}\left(\frac{H(\eta)}{w_s(\eta)}\right)~,
 \eeq
where we used $k_s=M_sa_s$ and $a_s/a=H/w_s$. The spectral tilt
of the axion perturbations is defined as $\gamma\equiv 3/2-r$.
If we define $\xi$ by $(\delta\rho_a)_k/\rho_a\propto k^\xi$, then the spectral tilt $\Delta n$ of the induced curvature
perturbations is given by $2\xi$, which in turn depends on $\gamma$ and the ratio $M_s/\psi_0$, as we will see in section III. For the Ramond-Ramond (R-R) axion
$m\neq 0$ unlike for the model independent axion, but in this
paper we assume $e^{\be_s}\sim 1$. We will alert the reader
whenever the physics is sensitive to this choice.

\section{Decaying axion as the origin of initial density
perturbations}

In this section we will estimate the spectral tilt of the density
fluctuations induced by the decay of the axion in the periodic
potential. We will assume that when the axion decays, it dominates
the energy density and thereby induces adiabatic initial
fluctuations for the CMB anisotropies. The periodicity of the
potential will damp the fluctuations which are larger than the
period of the potential. Due to this effect we are forced to
consider the case of a positive tilt spectrum $(\gamma>0)$ and
negative tilt spectrum $(\gamma<0)$ separately.

\subsection{Negative tilt fluctuations}

The relative density fluctuations of the axion field as the
potential is turned on is calculated in the Appendix. In case of
a negative tilted spectrum we arrive at
 \beq \label{qw}
\frac{\left(\delta\rho_a\right)_k}{\rho_a}
\approx\frac{1}{\sqrt{2}}\frac{\delta\psi_k}{\psi_0}e^{-\frac{1}
{2}\int_{k}^{k_s}\frac{dk}{k}\frac{\mathcal{P}_{\psi}(k)}{\psi^2_0}}~,
 \eeq
which agrees with the result in \cite{6}. Note that in our notation $\mathcal{P}_{\psi}=|\delta \psi|^2$. For the model
independent axion, which is the case $l=1$, the fluctuations will
be completely damped away by the exponential factor. But let us
examine the case with $l=0$ more carefully. Let us assume that
the spectral tilt $\gamma$ of the axion field fluctuations is
negative but very close to zero. In this case there is available
one additional degree of freedom, namely the string scale. To
evaluate the level of axion density fluctuations we split the
integral in Eq. (\ref{qw}) in two:
 \beq
\int_{k_A}^{k_s}\frac{dk}{k}\frac{\mathcal{P}_{\psi}(k)}{\psi_0^2}=\int_{k_A}^{k_*}\frac{dk}{k}
\frac{\mathcal{P}_{\psi}(k)}{\psi_0^2}+\int_{k_*}^{k_s}\frac{dk}{k}
\frac{\mathcal{P}_{\psi}(k)}{\psi_0^2}~,
 \eeq
where $k_A$ corresponds to the astrophysically interesting
comoving scale and $k_*$ is the comoving scale that has just
entered the horizon when axion starts to oscillate at
$\eta=\eta_*$. For the last term we get
 \beq
\int_{k_*}^{k_s}\frac{dk}{k}\frac{\mathcal{P}_{\psi}(k)}{\psi_0^2}=\frac{M_s^2}{\psi_0^2}\int_{k_*}^{k_s}
\frac{dk}{k}\left(\frac{k}{k_s}\right)^{2\gamma-2}\left(\frac{H(\eta_*)}{\omega_s(\eta_*)}\right)^2\approx
\frac{-1}{2\gamma-2}\frac{M_s^2}{\psi_0^2}\left(\frac{k_*}{k_s}\right)^{2\gamma}~
 \eeq
while
 \beq \label{koq}
\int_{k_A}^{k_*}\frac{dk}{k}\frac{\mathcal{P}_{\psi}(k)}{\psi_0^2}=\frac{1}{2\gamma}
\frac{M_s^2}{\psi_0^2}\left[
\left(\frac{k_*}{k_s}\right)^{2\gamma}-\left(\frac{k_A}{k_s}\right)^{2\gamma}\right]~.
 \eeq
In this way we obtain
 \beq \label{koq1}
\frac{\left(\delta\rho_a\right)_{k_A}}{\rho_a} =
\frac{M_s}{\psi_0}\left(\frac{k_A}{k_s}\right)^{\gamma}
e^{-\frac{1}{2}\frac{M_s^2}{\psi_0^2}\left[\frac{-1}{2(\gamma^2-\gamma)}\left(
\frac{k_*}{k_s}\right)^{2\gamma}-\frac{1}{2\gamma}\left(
\frac{k_A}{k_s}\right)^{2\gamma}\right]}~.
 \eeq
As we shall see in the next section, we may take $k_*/k_s\approx
10^{-7}$ within a few orders of magnitude. We did also use
$k_A/k_s\approx 10^{-30}$.

\begin{figure}[!hbtp] \label{kuva}
\begin{center}
\includegraphics[width=8cm]{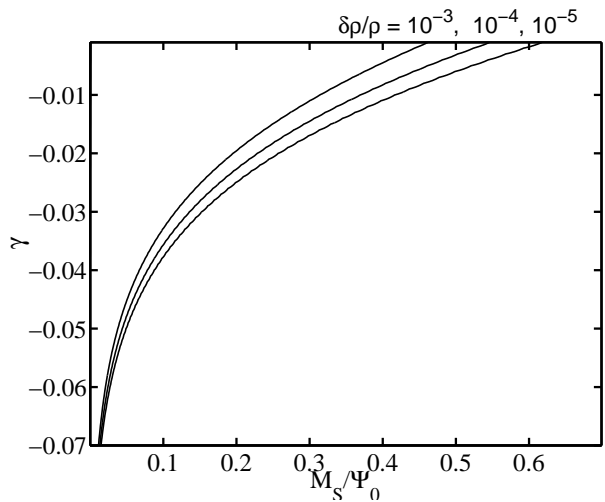}
\end{center}
\caption{Levels of constant $(\delta\rho_a)_k/\rho_a$ as a
function of $M_s/\psi_0$ on the horizontal axis and $\gamma$ on
the vertical axis. The three lines correspond to
$(\delta\rho_a)_k/\rho_a=10^{-3}$, $10^{-4}$, and $10^{-5}$.}
\end{figure}

In Fig. 1 we show a contour plot of
${\left(\delta\rho_a\right)_k}/{\rho_a}$ as given by Eq.
(\ref{koq1}) as a function of ${M_s}/{\psi_0}$ and $\gamma$. We
take the lower cut off on the momentum to be $k_{min}\simeq 0$ such that $\rho_a =V_0/2$ from (Eq.
(\ref{x})). For ${\left(\delta\rho_a\right)_k}/{\rho_a}=10^{-4}$
we find that the right level of density fluctuations are obtained
with $\gamma$ in a reasonable range $-0.04<\gamma< -10^{-5}$ when
$0.1<{M_s}/{\psi_0}<0.6$ (If ${M_s}/{\psi_0}\gtrsim 1$ then for a negative $\gamma$ the fluctuations are completely damped away for both $l=1$ and $l=0$).

From Fig. 1 we see that if $\psi_0$ differs only within a few orders of magnitude from $M_s$ we get $\gamma\gtrsim -0.05$ which correspond, as we shall see in the next section, to a bound on
the spectral tilt of the adiabatic density fluctuations $\Delta
n$ given by $0< \Delta n \lesssim 1$. This is potentially in
agreement with experimental bounds \cite{3}.

If we require $(\delta\rho_a)_k/\rho_a\sim 10^{-4}$ and set
$\gamma=-0.01$, we obtain $M_s/\psi_0\approx 0.4$. It is likely that string theory axions can have a decay constant
$\psi_0\approx 10^{17}\GeV$, which would lead to a present string
coupling in the lower limit of the theoretically favoured range
$g_s\sim 0.1 - 0.01$. In any case, the possibility of realizing
this idea depends very sensitively on the compactification.

When ${M_s}/{\psi_0}$ is very small, less than $10^{-4}$, the damping switches off. This means that even in this case $\gamma$ can be close to zero with a small but negative spectral tilt $\Delta n=2\xi=2\gamma$ (see next section). 

Finally we note that if $e^{\be_s}\neq 1$, the ratio $M_s/\psi_0$
either increases or decreases, depending on the sign of $m$. It is
interesting to note that $m$ can have both signs. In models with
an $SL(3,\mathbb{R})$-invariant effective action there are three
axions with respectively $m=0$, $m=-\sqrt{3}$, and $m=\sqrt{3}$
\cite{11}.

\subsection{Positive tilt fluctuations}
For the positive tilt case $(\gamma>0)$ one finds the following
spectrum (see the Appendix)
 \beq \label{q}
\frac{\left(\delta\rho_a\right)_k}{\rho_a} \approx
\frac{\delta\psi_k}{\sqrt{\left<\psi^2(\vec{x})\right>}}~.
 \eeq
At astrophysical scales we have $k_A/k_s\simeq 10^{-30}$ and at
these scales we want $(\delta\rho_a)_k/\rho_a\approx 10^{-4}$. It
it is not difficult to see that this is equivalent to requiring
 \beq
\left(\frac{k_A}{k_s}\right)^{\gamma}\approx 10^{-4}
 \eeq
which implies that the spectral tilt of the axion is
$\gamma \simeq 0.13$. This is much closer to flat than what is obtained
for adiabatic perturbations induced by the dilaton in the PBB
model but still slightly above the observational bounds, as we
will see shortly. However, in general we expect a combination of
isocurvature and adiabatic perturbations, which can change the
bounds somewhat.

The approximation used in the positive tilt case breaks down when
$(k_A/k_s)^{\gamma}\sim 1$, i.e. when $\gamma\sim 0$. Therefore
we can not exclude that a more careful analysis could show that
for $\gamma$ almost zero but positive one obtains also the right
level of density perturbations.

\subsection{Spectral tilt of the CMB anisotropies}

Let us now estimate the spectral index of the CMB fluctuations
corresponding to a slightly negative or positive value for the
axion tilt $\gamma$. We define $\Phi_k$ as in \cite{13}
 \beq
k^{-3/2}\frac{(\delta\rho)_k}{\rho}\equiv\delta_k=-\frac{2}{3}\left(\frac{k}{aH}\right)^2\Phi_k
 \eeq
whence one can calculate the relation of $\Phi_k$ to the
curvature fluctuation $\mathcal{R}_k$
 \beq
\Phi_k=-\frac{3+3w}{5+3w}\mathcal{R}_k~,
 \eeq
where $w=P/\rho$. Thus during radiation domination $w=1/3$.

The spectral index $n$ of adiabatic (curvature) fluctuations is
defined by
 \beq
\mathcal{P}_{\mathcal{R}}(k)\propto k^{n-1}
 \eeq
evaluated at the horizon entry; $n=1$ corresponds to the scale
invariant Harrison-Zel'dovich spectrum. Given that
 \beq
\frac{(\delta\rho)_k}{\rho}\propto k^{\xi}\qquad\Rightarrow
\delta_k=k^{\xi-3/2}
 \eeq
we get $n=2\xi+1$ so that the spectral tilt $\Delta n\equiv
n-1$ is given by
 \beq
\Delta n\equiv 2\xi
 \eeq

For a positive tilt with $\xi=\gamma\sim 0.13$ we would get a spectral
tilt $\Delta n=0.26$. The resulting index is slightly above
$n=1.06$ which is the upper limit given by Boomerang \cite{3},
although one should note that an uncorrelated isocurvature
component has the effect of raising the best fit for the
adiabatic spectral index \cite{14}. For a small negative
tilt $\gamma>-0.05$ the damping factor in eq.(\ref{qw}) implies
that this gives rise to a slightly positive tilted (blue)
spectrum of density fluctuations $0.3\lesssim \Delta n\lesssim
1$. Only if the ratio $M_s/\psi_0$ is very small, such that the damping switches off, it is possible to get a negative tilt. This result is obtained assuming that the axion density fluctuations are approximately unchanged by the coherent
oscillations and the subsequent decays. However if the amplitude
of the axion field fluctuations looses some magnitude as the
axions become non-relativistic it will make the spectrum, that
fits the amplitude at the COBE scale, much more flat. We will
return to this issue with a more detailed analysis in a later
paper.

Since the axion potential is highly non-linear, it is not trivial
to solve the perturbation equation mode-by-mode and to compute
the perturbations after the onset of the potential. For a
quadratic axion potential, valid for small field values, one
finds that at low frequencies the spectral tilt of the energy
density is unchanged after the onset of the potential \cite{10,6}.
However, the spectrum of a massive scalar field changes for modes
that become non-relativistic outside the horizon \cite{+,10}.
Therefore it would be important to study the evolution of the
field fluctuations on superhorizon scales after the potential is
generated and the axion mass has become larger than the Hubble
rate. In this way one could check whether the evolution of the
fluctuations on superhorizon scales affects the spectral tilt. In the present paper we have instead followed an approach
similar to that suggested in \cite{6,7}.

To understand what happens after the potential is turned
on, we divide the field $\psi$ in its large scale component,
which at scales $l=k^{-1}$ behaves as a constant classical field
$\psi_c(l^{-1})$, and its short wave length part $\delta\psi$
corresponding to momenta $k\geq l^{-1}$,
 \beq
\psi=\psi_c(l^{-1})+\delta\psi
 \eeq
where
 \beq
\psi_c(l^{-1})=\psi_c+\tilde\psi(l^{-1})~.
 \eeq
Here $\psi_c(l^{-1})$ includes the classical scale independent
field $\psi_c$ together with a contribution $\tilde\psi(l^{-1})$
from all the fluctuations with momenta smaller than $k$. The
dispersion of the long-wave perturbations
$\tilde\sigma_k=\sqrt{\sigma^2-\sigma_k^2}$, is given by
 \beq
\tilde\sigma^2_k=\int_{k_{min}}^{k}d\ln
k\left|\delta\psi(k)\right|^2~.
 \eeq
If $\tilde\sigma_k$ is comparable to $\psi_c$ then
$\psi_c(l^{-1})$ will indeed depend on $x$.

It is natural to assume that when the potential is turned
on, the field will oscillate a few times in the potential. As the
amplitude of the coherent oscillations decreases, the potential
effectively becomes quadratic. This is equivalent to the statement
that the axions will behave like non-relativistic matter when they
oscillate coherently in the potential. If $\psi_c(l^{-1})$ does
not change considerably at a distance $l\sim k^{-1}$ we find
\cite{7}
 \beq \label{eq1}
\frac{\rho_{\psi}(k)}{\rho_{\psi}}\sim \frac{\delta
V\left(\psi\left(k\right)\right)}{V\left(\psi\right)}\sim\frac{\delta\psi(k)}{\psi}
 \eeq
where $\psi$ can be identified with $\psi_c(l^{-1})$. This is
valid if $\psi_c(l^{-1})$ is larger than the short-wave length
dispersion $\sigma_k$. In the opposite limit we would get
 \beq \label{eq2}
\frac{\rho_{\psi}(k)}{\left<\rho_{\psi}\right>}\sim \frac{\delta
V\left(\psi\left(k\right)\right)}{\left<V\left(\psi\right)\right>}
\sim\frac{\left(\delta\psi(k)\right)^2}{\left<\right.\left(\delta\psi(k)\right)^2\left.\right>}~.
 \eeq
Since $\psi_c(l^{-1})$ evolves in the same way as $\delta\psi(k)$
and $\left(\delta\psi(k)\right)^2$ evolves in the same way as
$\left<\right.\left(\delta\psi(k)\right)^2\left.\right>$, we note
that the ratios in eq.(\ref{eq1}) and eq.(\ref{eq2}) remains
fixed up to some scale-independent factor that does not destroy
the scale invariance of the spectrum \cite{lyth}. This is
consistent with the calculation of \cite{10,6} where it was shown
for a quadratic potential that when the field becomes
non-relativistic the energy spectrum does not change on large
scales (up to a scale-independent factor).

In the case $\delta\rho_{\psi}\sim\delta\psi(k)$ in
eq.(\ref{eq1}) the density fluctuations will be gaussian, while in
the case of eq.(\ref{eq2}) the density fluctuations will have a
$\chi^2$-distributed non-gaussian nature \cite{lyth}. A $\chi^2$
perturbation is ruled out by observations \cite{obs}. Since large
field fluctuations are topologically cut off at the onset of the
potential, it seems likely that one will obtain non-gaussian
fluctuations unless the spectrum is very close to flat. Since the
evolution of $\delta\psi(k)$ is non linear from the moment when
the potential is turned on until the potential can be treated as
quadratic, a study of the extend of the non-gaussianity of the
density perturbations is beyond the scope of this paper.

\section{Reheat temperature and entropy production}
Let us now check that the axions have a chance of dominating the
energy density, as was assumed in the previous section. To
estimate the life time of the massive axions, we parameterize the
interaction between the axion and the gauge fields as
 \beq
(\psi/M')F\tilde{F}~,
 \eeq
where $M'\sim \pi^2\psi_0$ generally depends on the
compactification \cite{5,8}. A typical axion lifetime is
\cite{5,8}
 \beq
\tau_a\approx M'^2/m_a^3~,
 \eeq
where the axion mass $m_a$ is given by
 \beq
m_a\approx 10\frac{\Lambda^3}{M_pM'}~.
 \eeq
With $M'\sim 10^{18}\GeV$ we get $m_a=10^6 \GeV$ and thus
$k_*/k_s\sim 10^{-7}$ as claimed in the previous section. Like in
\cite{5} we will assume that $\Lambda\approx 10^{14}\GeV$.
Defining $R\equiv M'/M_p$ one finds that
 \beq
\tau_a\approx 10^{-3}R^5\frac{M_p^8}{\Lambda^9}\approx R^5\cdot
10^{23}\GeV^{-1}\approx \mathnormal{R}^5\cdot 10^{-2}~ \mathrm{sec}  ~.
 \eeq

The average energy density in the non-relativistic part of the
axion field after the potential is turned on reads
 \beq
\rho_a=\left<V(\psi(\vec{r}))\right>\approx \frac{1}{2}V_0~,
 \eeq
which implies that the relative energy density of the axions is
 \beq
\Omega_a=\frac{\rho_a}{\rho_c}=\frac{4\pi}{3}\frac{V_0}{M_p^2H^2}~.
 \eeq
To evaluate $\Omega_a$ at the time the potential is turned on, we
use the fact that at this time the Hubble paremeter $H$ has the
same order of magnitude as the axion mass. Typically we can also
take \cite{5,8}
 \beq
V_0\approx \frac{\Lambda^6}{M_p^2}
 \eeq
so that we obtain
 \beq
\Omega_a\approx 4\pi\frac{\Lambda^6}{M_p^4m_a^2}\simeq
10^{-1}\frac{M'^2}{M_p^2}=10^{-1}R^{2}~,
 \eeq
in agreement with \cite{15}. Thus the axions do not dominate
energy density at this point. This works only if we assume that
the axion fluctuations do not have a large negative tilt, which
would amplify small momentum mode fluctuations as they enter the
horizon and possibly make the axion dominate energy density
before the axion mass is turned on. If $\gamma>-0.05$ as in Fig.1
there is no problem.

The axion behaves as non-relativistic dust and soon starts to
dominate the energy density. Let us evaluate the reheat
temperature $T_{RH}$ as the axion decays into photons. Since
$H\approx 1.66 g_*^{1/2}T^2/M_{p}$ so that $H^2(t=\tau)\approx
\frac{1}{4}\tau^{-2}\approx 2.8g_* T_{RH}^4/M_p^2 $ which implies
that
 \beq T_{RH}= 0.55
g_*^{-1/4}\frac{M_p^{1/2}}{\tau^{1/2}}
 \eeq
we obtain
 \beq
T_{RH}\sim R^{-5/2}10^{-2}\GeV~.
 \eeq

In \cite{5} the entropy production associated with the decay of
the axion was calculated as
 \beq
\Delta s \simeq 10^{15} \left(
\frac{A_0}{M'}\right)^2\left(\frac{\Lambda}{10^{14}\GeV}\right)^{-3}R^4\Omega_r^{-3/4}
 \eeq
where $A_0\sim M'/\pi$ is the initial displacement of the axion
VEV and $\Omega_r\equiv N_r/N_{tot}$. $N_r$ is the number of
degrees of freedom in spin 0 and 1 fields charged under the gauge
groups of the observable sector. $N_{tot}$ is the total number of
degrees of freedom in spin 0, 1, and 2. We expect
$0.01\lesssim\Omega_r\lesssim 1$. By tuning the parameters
slightly, it is possible to obtain a reasonable amount of entropy
production of 8 to 10 orders of magnitude in order to dilute
dangerous relics \cite{5}.

If we define $T_*$ as the temperature of the universe when
$m_a=3H$, we find
 \beq \label{T*}
T_{*}\approx\sqrt{\frac{m_aM_p}{5g_*^{1/2}}} \approx
\sqrt{\frac{\Lambda^3}{RM_p}}\sim R^{-1/2}10^{11} \GeV~.
 \eeq
We denote the temperature as the axions decay by $T_D$. Noting
that $T_D<T_{RH}$ and $T_*>10^{11}\GeV$ we can conclude that
indeed the universe will become matter dominated soon after the
axion mass is turned on.

Let us choose as a typical example $R\sim 10^{-2}$. In this case
$T_{RH}=10^{3}\GeV$, but one needs to tune the other parameters in
order to have enough entropy production to dilute dangerous
relics. If $R\sim 1$, then one is about to get in trouble with a
too low reheat temperature $T_{RH}=10\MeV$, but there is plenty of
entropy production.

If baryogenesis is due to the decay of an Affleck-Dine (AD)
condensate, the AD condensate must not dominate energy density as
the axion decays. However it is possible that the decay of the AD
condensate produces a much too high baryon asymmetry and that the
late decay of the axion dilutes this to the level $n_B/s\sim (4
- 7)\times 10^{-11}$ \cite{5}.

\section{Conclusion}

We have shown that if one of the PBB axions decay to photons, it
is possible to get an adiabatic initial fluctuation for the CMB
anisotropies. Moreover, the considerations presented here seem to
point towards a small tilt, i.e. a nearly flat spectrum, although
the tilt of the initial adiabatic fluctuations may depend on the
axion oscillation and decay dynamics; here we have assumed that
the fluctuations are frozen during this time.

Our approach here is slightly different from \cite{5}, where it
was suggested that a negative tilt would make the axions dominate
energy density as the large wavelength modes enters the horizon,
and that the large fluctuations would be sufficiently damped by
the periodic potential. We have assumed that the spectrum is
nearly flat and the axions gets to dominate energy density
because they behave like non-relativistic dust when oscillating
in the harmonic potential. The fluctuations are damped, but the
relative magnitude of the string scale compared to the decay
constant plays a crucial role. In the positive tilt case we found
the additional possibility that the size of the tilt determines
the relative level of axion energy density fluctuations.

Since generically there is several string theory axions in the
type of models which we have discussed, it is possible that the
other axions, which are not expected to have decayed within the
present age of the universe, will carry isocurvature fluctuations
but with an amplitude smaller than the adiabatic fluctuations.
It seems thus viable that the scenario we have presented here will
generally lead to a mixture of adiabatic and isocurvature
fluctuations, with a dominating adiabatic component. Hence it appears that the
Pre-Big-Bang model does not necessaryly produce only initial
isocurvature fluctuations and can remain a potential candidate
for explaining the observed CMB data. Note also that the intermediate matter dominated phase could give a kink in the gravitational wave spectrum, which if observable would be an interesting feature.

A decaying axion may also be benefical from the point of view of
dangerous relics, which the entropy production associated with
the decay could dilute. A generic feature of this scenario is a
late reheat temperature which however is within the required
bound from Big Bang Nucleosynthesis.

We have kept our discussion general and did not constrain ourselves to any specific compactification scheme of the 10-dimensional string theory. It would however be important to investigate whether one can construct a specific model with the desired features.

\overskrift{Acknowledgements}

We would like to thank Riccardo Sturani for his interesting
comments about the gravitational wave spectrum. Martin S. Sloth
would also like to thank Fawad Hassan and Jussi V\"{a}liviita for
the inspiration from many discussions and especially Jussi for also lending out his matlab skills. K.E. partly supported by the Academy of Finland under the contracts 35224 and 47213. M.S.S. partly supported by the TMR network {\it Finite temperature phase transitions in particle physics}, EU contract no. FMRX-CT97-0122.

\appendix

\storoverskrift{Appendix}

\storoverskrift{Damping due to the periodic nature of the
potential}

Here we calculate the damping of the fluctuations in the
axion energy density. We present a simple intuitive argument.

Fluctuations in the axion energy density at scales $l\sim k^{-1}$
can be computed by using the following relation
 \beq
\int_0^{\infty}d\ln
k\left(\delta\rho_a^2\right)_k\frac{\sin(kr)}{kr}=\left<
V\left(\frac{\psi(\vec{x})}{\psi_0}\right)V\left(\frac{\psi(\vec{x}+\vec{r})}{\psi_0}\right)\right>-\left<
V\left(\frac{\psi(\vec{x})}{\psi_0}\right)\right>^2~.
 \eeq
By assuming that the fluctuatons in  $\psi$ are Gaussian such
that any connected correlation function of order higher than two
vanish, we get in agreement with \cite{6}
 \beq \label{hallo}
\int_0^{\infty}d\ln
k\left(\delta\rho_a^2\right)_k\frac{\sin(kr)}{kr}=\frac{1}{4}V_0^2e^{-\frac{\left<\psi^2(\vec{x})\right>}{\psi_0^2}}
\left[\cosh\left(\frac{\left<\psi(\vec{x})\psi(\vec{x}+\vec{r})\right>}{\psi^2_0}\right)-1\right]~.
 \eeq

To evaluate $(\delta\rho_a)_k$, we need to consider two limiting
cases
 \beq
\int_0^{1/l}d\ln
k\frac{\mathcal{P}_{\psi}(k)}{\psi^2_0}>>\int_{1/l}^{\infty}d\ln
k\frac{\mathcal{P}_{\psi}(k)}{\psi^2_0}
 \eeq
and
 \beq
\int_0^{1/l}d\ln
k\frac{\mathcal{P}_{\psi}(k)}{\psi^2_0}<<\int_{1/l}^{\infty}d\ln
k\frac{\mathcal{P}_{\psi}(k)}{\psi^2_0}
 \eeq
which respectively are typical for a negative and a positive
tilted spectrum.

Let us consider the first case. By integrating Eq. (\ref{hallo})
on both sides by $\int d^3x\int_{|\vec{k}|<1/l}d^3k
e^{i\vec{x}\cdot\vec{k}}$, we obtain
 \bea \label{hallo3}
\int_0^{1/l}d\ln k\left(\delta\rho_a^2\right)_k
=\frac{1}{4}V_0^2
e^{-\frac{\left<\psi^2(\vec{x})\right>}{\psi_0^2}} ~~~~~~~~~~~~~~~~~~~~~~~~~~~~
~~~~~~~~~~~~~~~~~~~~~~~~~~~~~~~~~~~~  &&\nonumber\\
\times \sum_{n=1}^{\infty}\frac{1}{2n!}\int\frac{d^3k_1}{4\pi
k_1^3}\frac{\mathcal{P}_{\psi}(k_1)}{\psi^2_0}\cdots
\int\frac{d^3k_{2n}}{4\pi
k_{2n}^3}\frac{\mathcal{P}_{\psi}(k_{2n})}{\psi^2_0}\int_{|\vec{k}|<1/l}d^3k\delta\left(\sum_{i=1}^{2n}
\vec{k}_i-\vec{k}\right)~.&&
 \eea
We first note that
 \beq
\int_{|\vec{k}|<1/l}d^3k\delta\left(\sum_{i=1}^{2n}
\vec{k}_i-\vec{k}\right)=\left\{\begin{array}{ccc}
  0 & if & |\sum_{i=1}^{2n}
\vec{k}_i|>1/l \\
  1 & if & |\sum_{i=1}^{2n}
\vec{k}_i|<1/l
\end{array}\right.
 \eeq
and assuming
 \beq\label{hallo1}
\int_0^{1/l}d\ln
k\frac{\mathcal{P}_{\psi}(k)}{\psi^2_0}>>\int_{1/l}^{\infty}d\ln
k\frac{\mathcal{P}_{\psi}(k)}{\psi^2_0}>0~,
 \eeq
we arrive at the following crude approximation
 \beq
\int_0^{1/l}d\ln k\left(\delta\rho_a^2\right)_k
\simeq\frac{1}{4}V_0^2
e^{-\frac{\left<\psi^2(\vec{x})\right>}{\psi_0^2}}
\sum_{n=1}^{\infty}\frac{1}{2n!}\int_0^{1/l}\frac{dk_1}{k_1}\frac{\mathcal{P}_{\psi}(k_1)}{\psi^2_0}\cdots
\int_0^{1/l}\frac{dk_{2n}}{k_{2n}}\frac{\mathcal{P}_{\psi}(k_{2n})}{\psi^2_0}~.
 \eeq
Instead of $0$ and $\infty$ as lower and upper limit for $k$, we
introduce the more physical upper and lower limits $k_s$ and
$k_{min}$. Then we get
 \beq
\int_{k_{min}}^{1/l}d\ln k\left(\delta\rho_a^2\right)_k
\approx\frac{1}{8}V_0^2e^{-\int_{1/l}^{k_s}\frac{dk}{k}\frac{\mathcal{P}_{\psi}(k)}{\psi^2_0}}
 \eeq
which agrees with the result in \cite{5}. If we use
$\left<\rho_a\right>=\frac{1}{2}V_0$ we obtain
 \beq
\frac{\left(\delta\rho_a\right)_k}{\rho_a}
\approx\frac{1}{\sqrt{2}}\frac{\delta\psi_k}{\psi_0}e^{-\frac{1}{2}\int_{k}^{k_s}\frac{dk}{k}\frac{\mathcal{P}_{\psi}(k)}{\psi^2_0}}~.
 \eeq
Hence in the negative tilt case the fluctuations are
exponentially damped, as discussed in section (3.1).

In the positive tilt case, we do not have the control of the
integral in Eq. (\ref{hallo3}) but calculate instead
 \bea \label{hallo4}
\int_{1/l}^{\infty}d\ln
k\left(\delta\rho_a^2\right)_k=\frac{1}{4}V_0^2e^{-\frac{\left<\psi^2(\vec{x})\right>}{\psi_0^2}}~~~~~~~~~~~~~~~~~~~~~~~~~~~~
~~~~~~~~~~~~~~~~~~~~~~~~~~~~~~~&&\nonumber\\
\times \sum_{n=1}^{\infty}\frac{1}{2n!}\int\frac{d^3k_1}{4\pi
k_1^3}\frac{\mathcal{P}_{\psi}(k_1)}{\psi^2_0}\cdots
\int\frac{d^3k_{2n}}{4\pi
k_{2n}^3}\frac{\mathcal{P}_{\psi}(k_{2n})}{\psi^2_0}\int_{|\vec{k}|>1/l}d^3k\delta\left(\sum_{i=1}^{2n}
\vec{k}_i-\vec{k}\right)~.&&
 \eea

In this case we use
 \beq
\int_{|\vec{k}|>1/l}d^3k\delta\left(\sum_{i=1}^{2n}
\vec{k}_i-\vec{k}\right)=\left\{\begin{array}{ccc}
  0 & if & |\sum_{i=1}^{2n}
\vec{k}_i|<1/l \\
  1 & if & |\sum_{i=1}^{2n}
\vec{k}_i|>1/l
\end{array}\right.
 \eeq
which motivates the following approximation
 \bea
\int_{1/l}^{\infty}d\ln k\left(\delta\rho_a^2\right)_k
=\frac{1}{4}V_0^2e^{-\frac{\left<\psi^2(\vec{x})\right>}{\psi_0^2}}~~~~~
~~~~~~~~~~~~~~~~~~~~~~~~~~~~~~~~~~~~~~~~~~~~~~~~~~~~~~~~~~&&\\
\times
\sum_{n=1}^{\infty}\frac{1}{2n!}\int_{0}^{\infty}\frac{dk_1}{k_1}\frac{\mathcal{P}_{\psi}(k_1)}{\psi^2_0}\cdots
\int_{0}^{\infty}\frac{dk_{2n-1}}{
k_{2n-1}}\frac{\mathcal{P}_{\psi}(k_{2n-1})}{\psi^2_0}\int_{1/l}^{\infty}\frac{dk_{2n}}{
k_{2n}}\frac{\mathcal{P}_{\psi}(k_{2n})}{\psi^2_0}~.&&
 \eea
Again we introduce the more physical upper and lower limits $k_s$
and $k_{min}$ and obtain
 \beq
\frac{\left(\delta\rho_a^2\right)_k}{\rho_a^2}
\approx\frac{1}{2}\frac{\delta\psi_k^2}{\left<\psi^2(\vec{x})\right>}\left(
\frac{1-e^{-\frac{\left<\psi^2(\vec{x})\right>}{\psi_0^2}}}{1-e^{-\frac{1}{2}
\frac{\left<\psi^2(\vec{x})\right>}{\psi_0^2}}} \right)^2\approx
\frac{\delta\psi_k^2}{\left<\psi^2(\vec{x})\right>} ~.
 \eeq
As discussed in section (3.2), we find for the positive tilt case
that the fluctuations are not damped.

\end{document}